\documentclass[a4paper,twocolumn,11pt,accepted=2023-07-11]{quantumarticle}
\pdfoutput=1
\usepackage[utf8]{inputenc}
\usepackage[english]{babel}
\usepackage[T1]{fontenc}
\usepackage{amsmath}
\usepackage{hyperref}

\usepackage{amssymb}
\usepackage{graphicx}
\usepackage{caption}
\usepackage{subcaption}

\usepackage{tikz}
\usepackage{lipsum}

\usepackage[numbers]{natbib}
\usepackage{ulem}

\begin{document}

\title{Chiral superconductivity in the doped triangular-lattice Fermi-Hubbard model in two dimensions}
\author{Vinicius Zampronio}
\affiliation{Institute for Theoretical Physics, Utrecht University, 3584CS Utrecht, Netherlands}
\affiliation{Departamento de F\'isica Te\'orica e Experimental, Federal University of Rio Grande do Norte  59078-950 Natal-RN, Brazil}
\author{Tommaso Macrì}
\affiliation{ITAMP, Harvard-Smithsonian Center for Astrophysics, Cambridge, Massachusetts 02138, USA}
\affiliation{Departamento de F\'isica Te\'orica e Experimental, Federal University of Rio Grande do Norte  59078-950 Natal-RN, Brazil}
\maketitle

\begin{abstract}
    The triangular-lattice Fermi-Hubbard model has been extensively investigated in the literature due to its connection to chiral spin states and unconventional superconductivity.
    Previous simulations of the ground state of the doped system rely on quasi-one-dimensional lattices where true long-range order is forbidden.
    Here we simulate two-dimensional and quasi-one-dimensional triangular lattices using state-of-the-art Auxiliary-Field Quantum Monte Carlo. Upon doping a non-magnetic chiral spin state, we observe evidence of chiral superconductivity supported by long-range order in Cooper-pair correlation and a finite value of the chiral order parameter. With this aim, we first locate the transition from the metallic to the non-magnetic insulating phase and the onset of magnetic order. Our results pave the way towards a better understanding of strongly correlated lattice systems with magnetic frustration.
\end{abstract}

As a paradigmatic model for strongly correlated fermionic lattice systems,
the Fermi-Hubbard (FH) Hamiltonian still has many open questions~\cite{aro22}.
In two dimensions (2D), the FH model captures the rich physics of metal to insulator phase transitions (MIT)~\cite{ima98},
itinerant magnetism~\cite{hir85} and spin liquids~\cite{bal10,sav16,zho17}.
Quantum spin liquids with non-abelian anyon excitations can act as building blocks for topological quantum computation and for the construction of fault-tolerant quantum computers~\cite{che08}.
The idea behind this non-magnetic insulator was proposed in the seminal work by Anderson that describes
a resonating valence-bond state arising from geometric frustration on the lattice~\cite{and73,and87}.
The simplest lattice structure containing this kind of frustration is the triangular one,
which is relevant for the understanding of molecular materials of the $\kappa$-ET family~\cite{shi03,kur05,yam08,iso14,mik21}.
Besides geometrical frustration,
charge fluctuations play a role in the stabilization of quantum spin liquids~\cite{mot05,lee05}.
The simulation of charge fluctuations can be accomplished by introducing high-order ring-exchange coupling
to the effective spin Hamiltonian~\cite{sch07,she09,yan10,coo21}
or by considering the Hubbard model itself.
Additionally,
the observation of Hubbard-model physics in triangular lattices engineered in WeSe$_2$/WeS$_2$ moir\'{e} superlattices~\cite{wu18,tan20}
and in quantum simulators~\cite{yang21,mon22} has inspired scientific interest in such systems.
In this work, we focus on the triangular-lattice FH model described by the Hamiltonian
\begin{equation}\label{eq_fhh}
H=-t\displaystyle\sum_{\left< ij\right>,\alpha}\left(c^{\dagger}_{i\alpha}c_{j\alpha}+\text{H.c.}\right)
+U\displaystyle\sum_i n_{i\uparrow}n_{i\downarrow},
\end{equation}
where $\alpha=\uparrow,\downarrow$ is the electron spin, $\left< ij\right>$ indicates a sum over nearest-neighbour sites, $c_{i\alpha}$ and $c^{\dagger}_{i\alpha}$ respectively annihilates and creates an electron with spin $\alpha$ at the $i$-th lattice site, $n_{i\alpha}= c^{\dagger}_{i\alpha}c_{i\alpha}$ is the number operator, $t$ is the hopping strength, and $U$ is the intensity of the onsite interaction.
The non-interacting system is metallic and,
at half-filling $(\langle n_{i\alpha} \rangle =1/2)$,
the strongly interacting FH Hamiltonian can be mapped into the antiferrognetic (AFM) Heisenberg one, whose ground state contains $120^{\circ}$ long-range spin order~\cite{whi07}.
Away from these two cases ($U/t=0$ and $U/t\gg1$),
the precise nature of the system is still under debate.
For weak interactions,
numerical simulations assume an adiabatic connection with the non-interacting regime.
Nonetheless, they might be overlooking a transition to a phase with a small but non-vanishing gap~\cite{aro22}. In fact, renormalization-group calculations predict a $d+id$  superconductor at $U/t \ll 1$ in 2D~\cite{rag10,rah14} and weak coupling analyses argue that, at weak interactions, the quasi-1D system is a Luther-Emery liquid with time-reversal symmetry breaking~\cite{gan20}.  
For intermediate interactions,
the quest for spin liquids has been a subject of significant interest.
Still,
there is not even theoretical agreement on whether the spin liquid state exists in the FH model.
While calculations ranging from variational cluster approximation~\cite{sah08,yam14,lau15},
path integral renormalization group~\cite{mor02,yos09},
strong coupling expansion~\cite{yan10},
dual fermion approach~\cite{ant11} and
exact diagonalization~\cite{kor07}
to density matrix renormalization group (DMRG)~\cite{shi17,sza20,sza21,che22} and variational Monte Carlo (VMC)~\cite{toc20,toc21} agreed in the existence of a spin liquid state,
dynamical cluster approximation studies~\cite{lee08} and earlier VMC computations~\cite{wat08,toc13} detected a direct transition from a metallic state to a magnetic ordered phase.
Among the theories that support the existence of a spin liquid, its nature remains controversial.
Infinite-DMRG calculations predict a gapped chiral spin liquid (CSL)~\cite{sza20,sza21},
while VMC simulations on full 2D systems and  finite-DMRG~\cite{shi17} support a gapless spin liquid that preserves time-reversal symmetry.
Another finite-DMRG study also supports the gapped CSL~\cite{che22}.
On the other hand, a multi-method approach finds that, at intermediate interactions, there is a competition between chiral and two distinct magnetic orders: collinear and $120^{\circ}$ order~\cite{wie21}.
DMRG simulations of the extended AFM-Heisenberg model with four-spin interactions that arise naturally from Mott-insulator physics
corroborate the existence of a CSL in lattice geometries closer to 2D than the ones used in DMRG simulations of the Hubbard model~\cite{coo21}.
For the hole-doped system,
the quest for unconventional superconductivity in the Hubbard model is a matter of current scientific interest due to its connection to High-$T_c$ superconductors~\cite{lee06,pow11,kan11,boh21-2}.
A recent DMRG study of the doped triangular FH predicts a rich phase diagram with fractionalized  excitations, spin and charge deconfinement and enhanced Cooper-pair correlations~\cite{zhu22}.
Another DMRG study estimates the spectral function of one single hole doped in the triangular-lattice CSL and observes spinon dynamics~\cite{kad22}.
Also, DMRG simulations of the extended $t-J$ model with three-spin chiral interactions in the triangular lattice predicted chiral superconductivity in the system, evidenced by quasi-long-range-order in the Cooper-pair correlations upon doping~\cite{hua22}.
Shortly thereafter, emergent topological superconductivity was also reported in the simpler $t-J$ model~\cite{hua23}.
However, DMRG performed in quasi-$1$D lattices does not display  true long-range order, and one has to rely on a slow decay of Cooper-pair correlations.
Numerical simulations via the Linked-Cluster Expansion algorithm provide several benchmarks to the finite temperature triangular FH at intermediate to strong interactions~\cite{gar22},
but a clear description of the weakly interacting regime, the classification of the spin liquid state, and whether or not a superconducting phase would appear upon doping is still elusive.

In this work, we numerically investigate the ground state of the doped triangular FH in 2D.
Upon doping a non-magnetic chiral spin state (CSS) we observe true long-range order in the Cooper-pair correlations while the chiral order parameter remains finite, i.e. a chiral superconductor.
To simulate the CSS, we first locate the MIT and the transition to the AFM phase.

\section{Methods}

We report the implementation of state-of-the-art Auxiliary-Field Quantum Monte Carlo (AFQMC) to simulate the ground state of the full 2D triangular lattice FH model.
By imaginary-time projection to the ground state we intend to reduce the bias from VMC calculations. 
A constrained-path (CP) approximation is required to restore polynomial convergence (otherwise plagued by the sign problem ~\cite{tro05}) and the bias from the variational ansatz is not completely removed. However, simulations made in the past for square lattices away from half-filling have been shown to be accurate and provided several benchmarks~\cite{zha95,zha97,ngu14,leb15,qin16}. 
See Appendix~\ref{sec_mc} for further details of the method and for a comparison of the CP-AFQMC estimates of the triangular-lattice ground-state energy with exact diagonalization.

For the non-magnetic phases, we consider the generalized Hartree-Fock ansatz (GHF) for the imaginary-time projection.
The mean-field Hamiltonian associated to the GHF state is obtained considering a partial particle-hole transformation on a BCS Hamiltonian,
\begin{equation}\label{eq:mfh}
H_{\text{MF}} = -t\sum_{\langle i j\rangle \alpha} c^{\dagger}_{i \alpha}c_{j\alpha}+ \sum_i M_i c^{\dagger}_{i\uparrow}c_{i\downarrow} + \text{H.c.},
\end{equation}
where $M_i=U_{\text{eff}}\langle c^{\dagger}_{i\uparrow}c_{i\downarrow} \rangle$.
The GHF ground state is obtained via a self-consistent diagonalization of Eq.~\eqref{eq:mfh}~\cite{qin16}. As done with Unrestricted Hartree Fock (UHF) wave functions~\cite{leb15}, we consider $U_{\text{eff}}=\min(U,U_{\text{eff}}^{\text{max}})$.
We noticed that $U_{\text{eff}}^{\text{max}}/t=4$ gives  meaningful estimates of the non-magnetic states of the system.
Our CP-AFQMC simulations with the GHF ansatz became unstable for strong interactions, see Appendix~\ref{sec_mc} for more explanation concerning the instability. With this ansatz, we were able to locate the MIT but not the transition to the magnetic phase.
To see the transition to the $120^{\circ}$ AFM phase, we also perform simulations starting from a full Hartree-Fock (FHF) ansatz. The ansatz that provides the smaller energy after imaginary-time evolution is our best representative of the system ground state, see Appendix~\ref{sec:mf} for details.

\begin{figure}[h]
\centering
\includegraphics[scale=0.4]{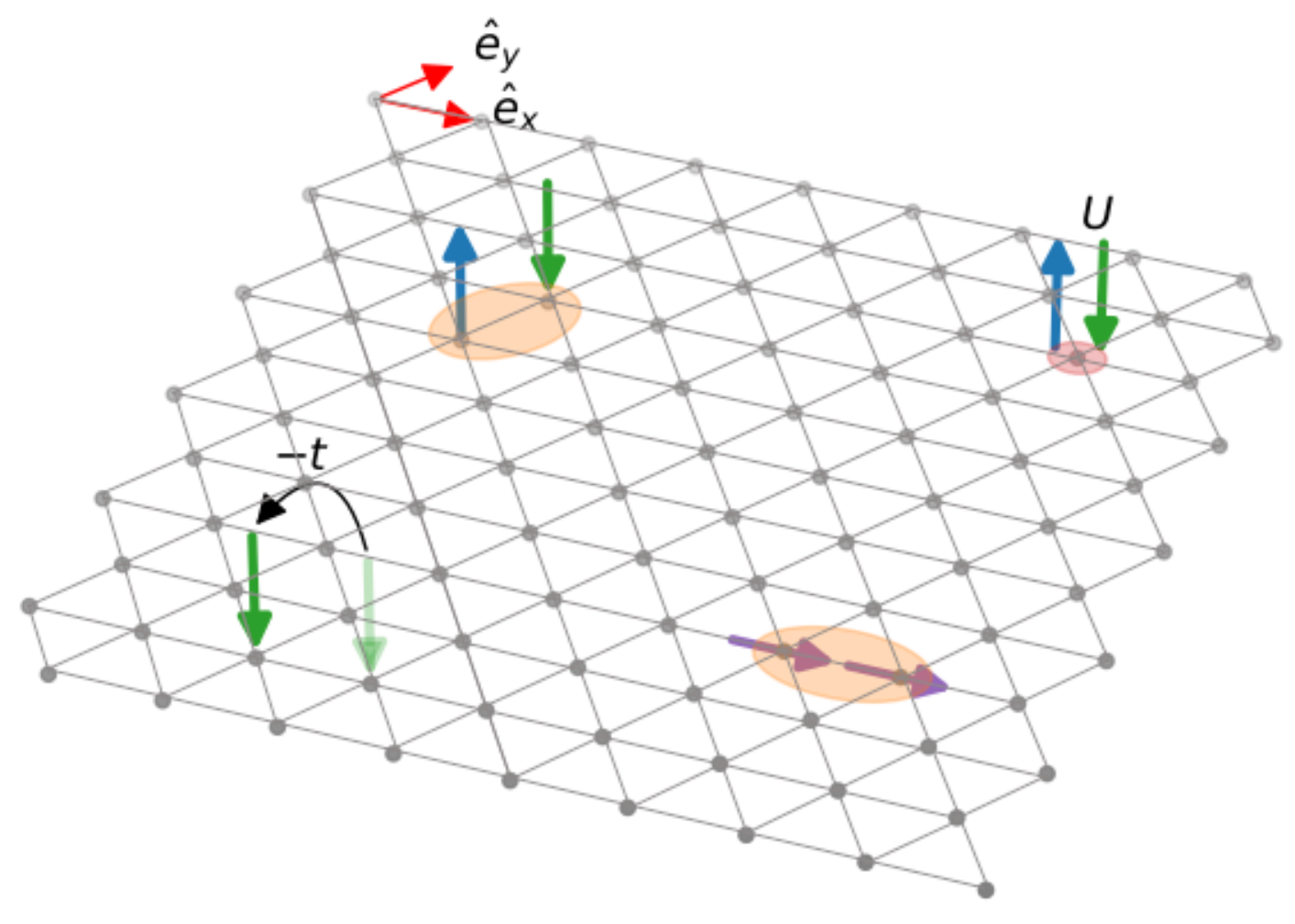}
\caption{{\bf Triangular FH model.} We display the lattice vectors $\hat{e}_x = (a,0)$ and $\hat{e}_y=(a/2,\sqrt{3}a/2)$.
The circle depicts the formation of a doublon (doubly-occupied site) with energy cost $U$. 
We also represent a hopping process with energy scale $-t$.
Ellipses depict Cooper pairs in the singlet ($|s\rangle = (|\uparrow\downarrow\rangle - |\downarrow\uparrow\rangle) /\sqrt{2}$) or triplet ($|t\rangle = (|\uparrow\downarrow\rangle + |\downarrow\uparrow\rangle) /\sqrt{2}$) states formed on the $\hat{e}_y$ and $\hat{e}_x$ bounds of the triangular plaquettes.
\label{fig:tl}}
\end{figure}

In our simulations, we mainly consider lattices with $N_x = N_y = 12$ sites along the $\hat{e}_x = (a,0)$ and $\hat{e}_y=(a/2,\sqrt{3}a/2)$ directions respectively (see FIG.~\ref{fig:tl} for the lattice vectors and a visual representation of the triangular FH model).
Periodic boundary conditions are considered along the horizontal and vertical directions.
We also run simulations with different lattice sizes to  analyse the finite-size effects.
See the Appendix~\ref{sec:finsize} for the dependence of total energy and spin correlations on the lattice size.
Finally, to address the effect of the dimensionality, we investigate quasi-1D lattices with $N_x = 36$, $N_y = 3$ and $4$, PBC along $\hat{e}_y$ and open boundary condition along $\hat{e}_x$.
From now on, if not specified otherwise, we are considering the $12\times 12$ lattice.
We study spin-balanced systems at half filling ($n=N/M=1$, where $N$ is the number of electrons and $M=N_xN_y$) and with hole doping ($n<1$).

\section{Results at half filling}

\begin{figure}[h]
\centering
\includegraphics[scale=0.85]{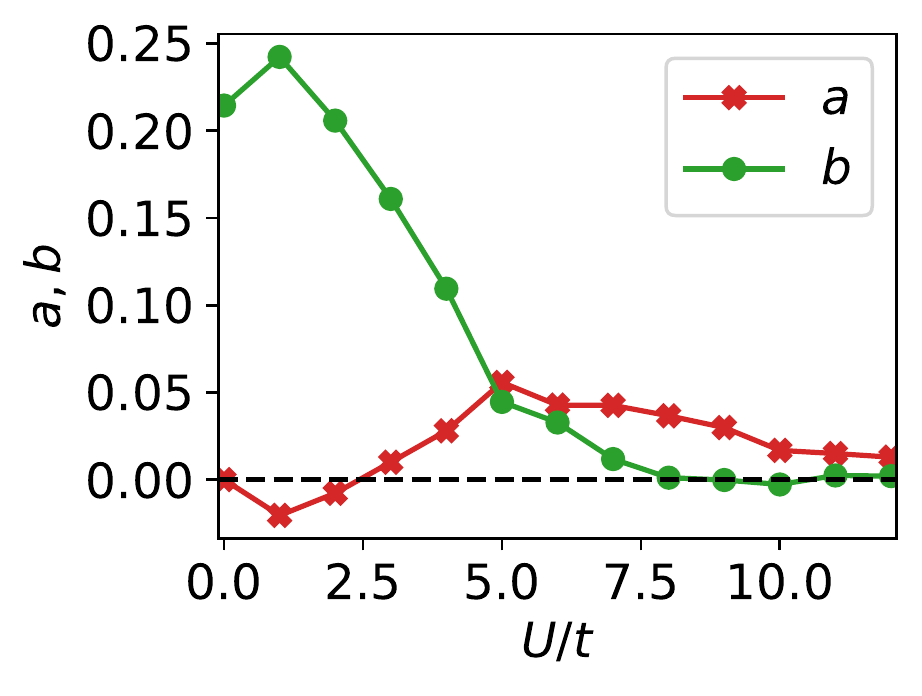}
\caption{{\bf Metal-insulator transition}.
The coefficients of the expansion $N(k)\approx ak^2 + bk$ as a function of the interaction $U/t$.
A nonzero value of $b$ implies that $\Delta_c = 0$.
\label{fig:mit}}
\end{figure}

We start by defining the charge structure factor
\begin{equation}\label{eq:nk}
    N(\mathbf{k}) = \frac{1}{M}\sum_{i,j}
    e^{i\mathbf{k}\cdot\mathbf{r}_{ij}}
    \left(\langle n_in_j\rangle - \langle n_i \rangle\langle n_j\rangle
    \right),
\end{equation}
with $n_{i}=n_{i \uparrow}+ n_{i \downarrow}$,
around the origin $\mathbf{k} = 0$ to determine whether the system is gapped or not.
Since the charge gap $\Delta_c$ is proportional to $\lim_{k \rightarrow 0}k^2/N(k)$~\cite{fey54,cap05},
a linear behavior around $k=0$ indicates that the system is metallic (gapless) and
a quadratic behavior indicates that the system is insulating (gapped).
More precisely,
for small $k$ we have $N(k) = ak^2 + bk +\mathcal{O}(k^3)$,
and whenever $b \neq 0$, $\Delta_c = 0$.
We compute the coefficients $a$ and $b$ by performing a quadratic fit to our data.
We considered the three allowed momenta closer to the origin along $\mathbf{k}=(k_x=0,k_y)$,
results are shown in FIG.~\ref{fig:mit} where we located the MIT around $7 < U_{c1}/t \leq 8$.
Our critical interaction is in agreement with DMRG simulations~\cite{shi17,sza20} and
a multi-method study~\cite{wie21}.
The Linked-Cluster-Expansion calculations, after an extrapolation to zero-temperature regime, predict a critical interaction $U_{c1}/t \sim 7$~\cite{gar22},
while other DMRG simulations predict $U_{c1}/t \sim 9$~\cite{che22}.

\begin{figure}[h]
\centering
\includegraphics[scale=0.85]{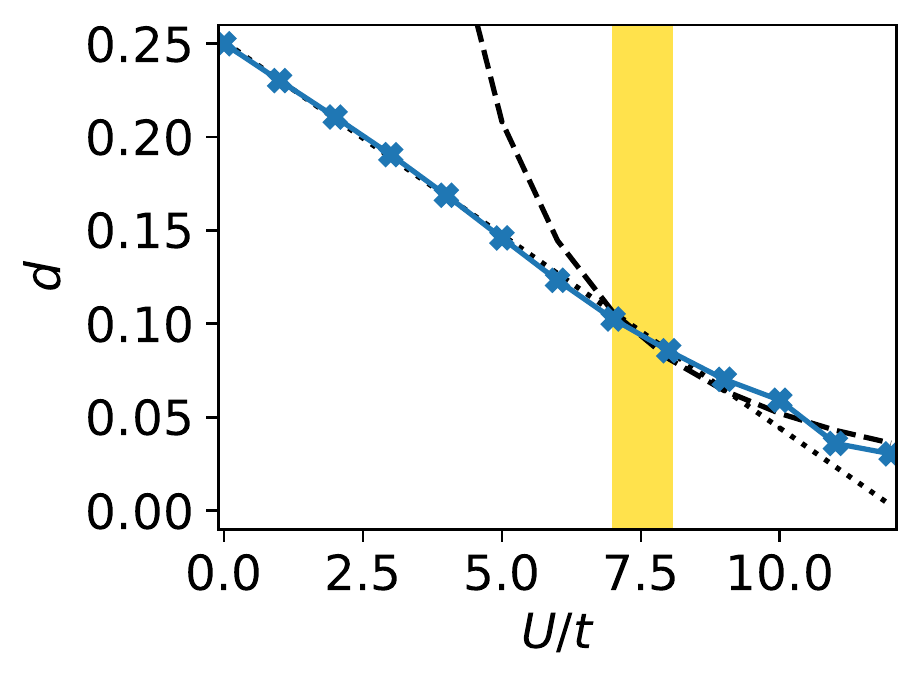}
\caption{{\bf Doublon density} as function of the interaction strength $U/t$. The blue crosses represent our data. The dotted line is a linear fit of the doublon-density data for $U/t < 7$. The shaded area delimits the region where $U_{c1}$ is located. The dashed curve is the function $d = (2t^2/U^2\sum_{\delta}(1/4-\langle \mathbf{S}_i\cdot\mathbf{S}_{i+\delta}\rangle)$
with $\langle \mathbf{S}_i\cdot\mathbf{S}_{i+\delta}\rangle = -0.1837(7)$, see Ref.~\cite{li15}.
Error bars are smaller than the markers size.\label{fig_ddens}}
\end{figure}

To further corroborate our findings, we compute the doublon density,
$d =\sum_i\langle n_{i\uparrow}n_{i\downarrow} \rangle/M$,
for which we expect different behaviors in the metallic and insulating phases~\cite{kok12,shi17}.
In the metallic phase the Brinkman-Rice picture predicts that $d$ decreases linearly with $U$~\cite{bri70},
while for strong interactions the doublon density shows Heisenberg behavior~\cite{esk94},
$d = (2t^2/U^2\sum_{\delta}(1/4-\langle \mathbf{S}_i\cdot\mathbf{S}_{i+\delta}\rangle )$,
where the sum runs over the nearest neighbours.
In Fig.~\ref{fig_ddens} we display results for $d$ as a function of $U/t$.
Our data for the doublon density show a deviation from the linear behavior near the MIT.

\begin{figure}[h]
\centering
\includegraphics[scale=0.85]{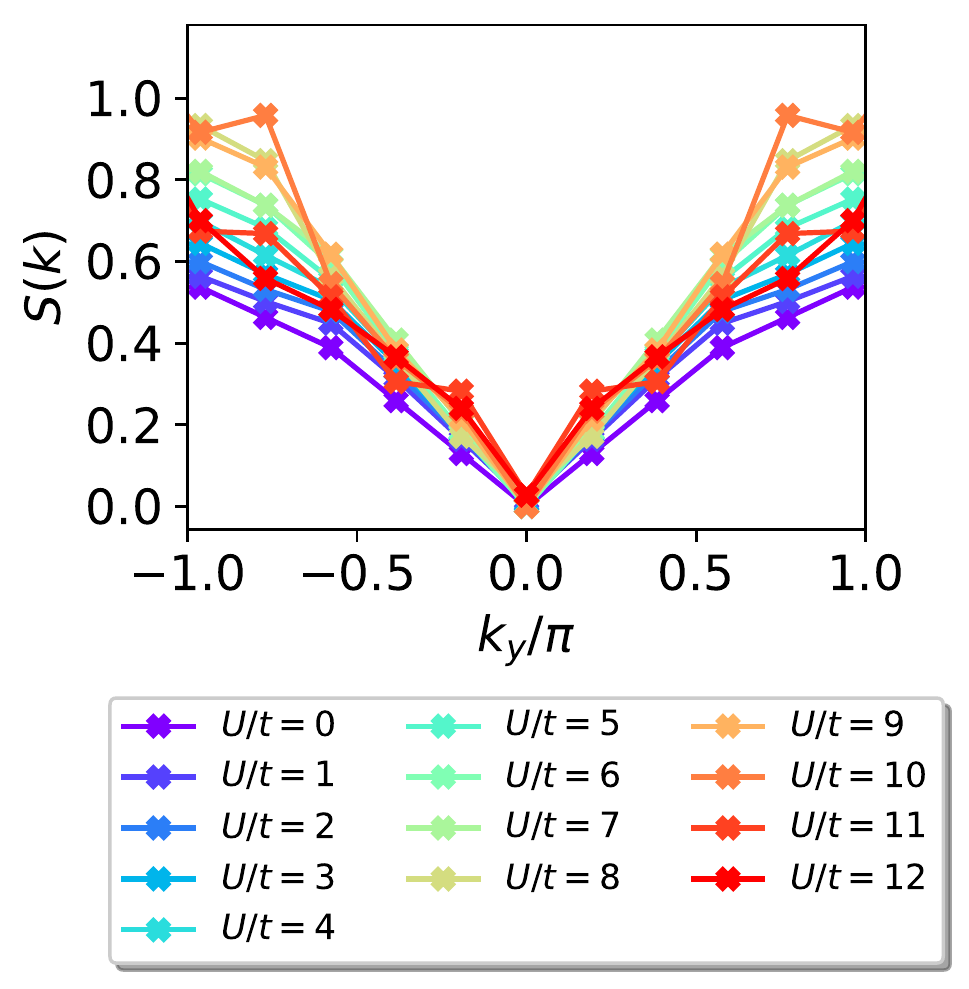}
\caption{{\bf Spin structure factor} $S(k)$ as a function of $U/t$ for $\mathbf{k}=(k_x=0,k_y)$. The linear behavior around $k_y=0$ indicates gapless spin excitation.\label{fig:sk}}
\end{figure}

Analogously, the presence of a spin gap can be accessed by the spin structure factor
\begin{equation}
S(\mathbf{k}) = \frac{1}{M} \sum_{i,j}e^{i\mathbf{k}\cdot\mathbf{r}_{ij}}\langle S^z_iS^z_j \rangle,
\end{equation}
with $S^z_{i}=n_{i\uparrow}-n_{i\downarrow}$.
We do not see the emergence of quadratic behavior in $S(k)$ (Fig.~\ref{fig:sk}),
which indicates the absence of a spin gap.
The excitations of the 120$^{\circ}$ Heisenberg antiferromagnet are gapless magnons~\cite{che09}.
Therefore the presence of a spin gap is not a good measure to locate the AFM order in our system.
On the other hand, the presence of peaks in $S(k)$ is a signature of long-range magnetic order; for the $120^{\circ}$ AFM those peaks appear on the $K$ points of the Brillouin zone.
In our simulations, we see the formation of peaks on the $K$ points which indicates the transition to the 120 $^{\circ}$ phase with critical interaction $10 < U_{c2}/t \le 11$ (see Appendix~\ref{sec:1-2D}). Our estimate for the critical interaction is in agreement with recent DMRG simulations~\cite{sza20}, which locates the transition at $U/t = 10.6$.

We investigate the chiral order parameter

\begin{equation}\label{eq:chi}
    \chi = \left|\sum_{\triangle} \langle \mathbf{S}_i\cdot(\mathbf{S}_j\times\mathbf{S}_k)\rangle \right|,
\end{equation}
where the sum is over every triangular plaquette of the lattice with vertexes $i$, $j$ and $k$ taken in the clockwise direction and
$\mathbf{S}_{\ell}=\sum_{\alpha,\beta=\uparrow,\downarrow}c^{\dagger}_{\ell\alpha}
\mathbf{\sigma}_{\alpha\beta}c_{\ell\beta}$,
with $\mathbf{\sigma}=(\sigma^x,\sigma^y,\sigma^z)$ a vector of Pauli matrices.

\begin{figure}[h]
\centering
\includegraphics[scale=0.75]{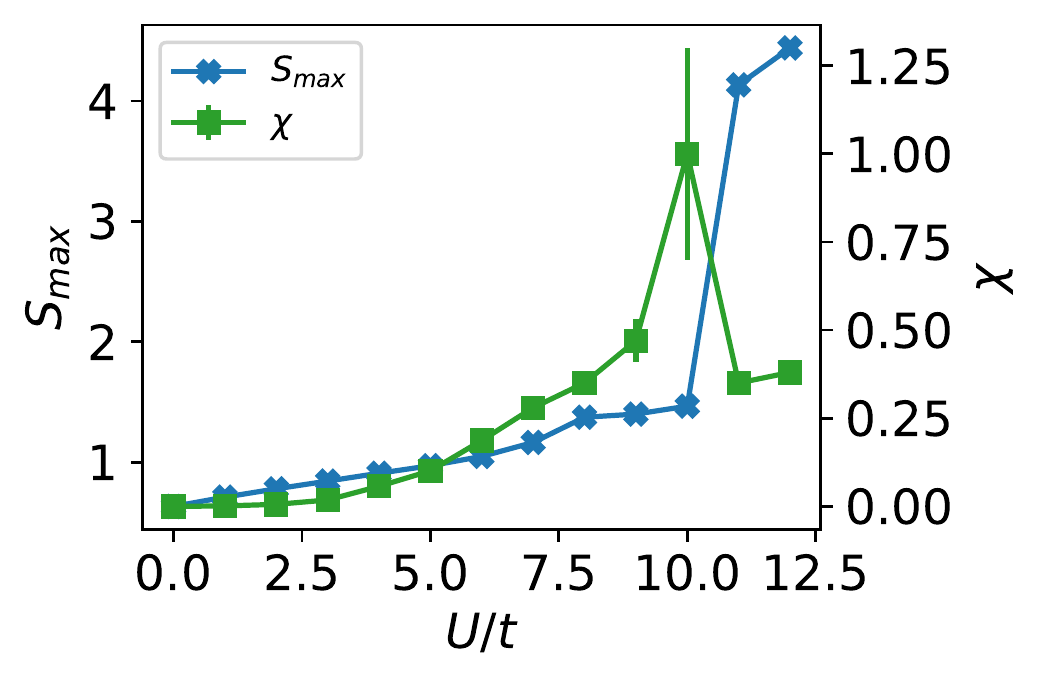}
\caption{{\bf Magnetic versus chiral orders}. Maximum value of S(k) (crosses). Chiral order parameter $\chi$ (squares). The chiral parameter vertical axis is on the right. \label{fig:magvschi}}
\end{figure}

Our results for $\chi$ are show in Fig.~\ref{fig:magvschi} alongside the $S_{\text{max}}$ of $S(k)$. We observe a competition between chiral and magnetic orders as the interaction increases as reported for quasi-1D lattices~\cite{wie21}.
We see a sharp increase in $S_{\text{max}}$ in the transition to the AFM phase.

To analyse the effect of dimension in our results, we compute $S(k)$ and $\chi$ for the quasi-1D $36\times 3$ lattice. We see that the results in 2D and quasi-1D agree reasonably well, but in quasi-1D we see the AFM insulator with the peaks of the spin structure factor at the $M$ points, which is a signature of a collinear AFM phase. We also analyse the effect of the width of the quasi-1D systems on the results by the simulation of a $36\times 4$ lattice, where we see competition between $120^{\circ}$ and collinear orders. See Appendix~\ref{sec:1-2D} for the results of the quasi-1D simulations. The existence of a CSS was predicted in the non-magnetic insulating phase by DMRG~\cite{sza20,che22,zhu22}. We also detect finite chiral order before the transition to the insulating phase ($U/t \leq 7$).
This metallic chiral state is in disagreement with the predictions of Ref.~\cite{sza20}, where the authors perform infinite-DMRG and extrapolate the chiral-order parameter $\chi$ on the bond dimension. A extrapolation of this sort is not needed in CP-AFQMC simulations since the summation over auxiliary-field paths restores all the correlations in the system. Moreover, other DMRG simulations of this system at weak doping also report the observation of a chiral metal~\cite{zhu22}. Another difference, between our results and the ones of Ref.~\cite{sza20} after extrapolation, is the finite value of $\chi$ near the transition to the magnetic phase ($11 \leq U/t \leq 12$).Nevertheless, the computation of the chiral susceptibility in Ref.~\cite{wie21} also shows evidence of chirality in the magnetic phase.

\section{Doping the non-magnetic CSS}

\begin{figure}[h]
\centering
\includegraphics[scale=0.65]{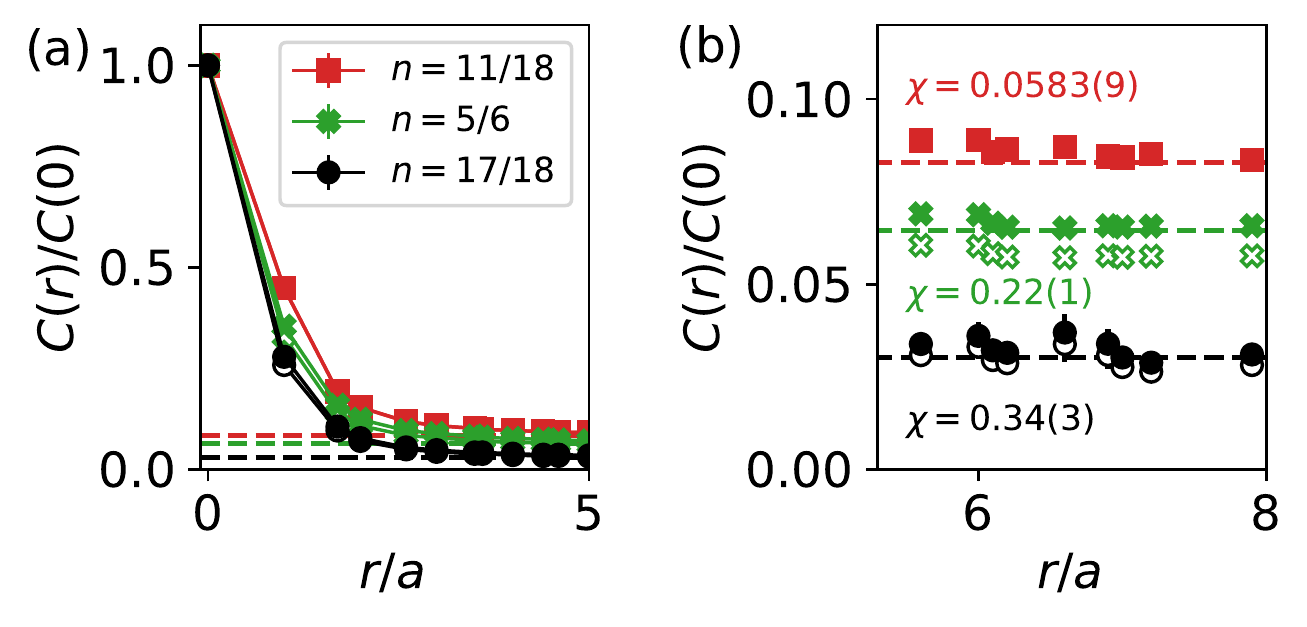}
\caption{{\bf Cooper pair correlations}.
(a) Filled (empty) markers represent correlations between pairs in the triplet (singlet) state.
Dashed lines represent an average of the correlations for $r/a \gtrsim 5$ (colors are matching).
(b) Long-range behavior of the correlations,
the chiral order parameter $\chi$ of each system is shown (colors are matching).
The interaction strength is $U/t=9$. When not shown, error bars are smaller than the markers size.
\label{fig:cooper}}
\end{figure}

We search for signals of superconductivity in the doped system by looking at the long-range behavior of Cooper-pair correlations.
Following the procedure described in Ref.~\cite{zhu22} we define the correlation function
\begin{equation}\label{eq:cooper}
    C(\mathbf{r})=\frac{1}{3M}\sum_{i,\hat{\delta}}\left|\langle\Delta^{\dagger}_{i,\hat{e}_y}\Delta_{i+\mathbf{r},\hat{\delta}}\rangle \right|,
\end{equation}
where $\hat{\delta}= \{\hat{e}_x, \hat{e}_y, \hat{e}_x-\hat{e}_y\}$.
The superconductivity order parameter $\Delta$ is given by
$\Delta_{\ell,\hat{\delta}}=(1/\sqrt{2})\sum_{\alpha}f(\alpha)c_{\ell\alpha}c_{\ell+\hat{\delta}\bar{\alpha}}$, where $\alpha = \uparrow,\downarrow$ and $\bar{\alpha}$ is the opposite spin of $\alpha$. For pairs in the singlet state, $f(\uparrow)=-f(\downarrow)=1$, and for triplets $f(\uparrow)=f(\downarrow)=1$.
We compute angle-averaged $C(r)$ in the non-magnetic CSS ($U/t=9$) for three distinct fillings of the hole doped system,
$n = 17/18,$ $5/6$ and $11/18$, covering the weak-, intermediate-, and strong-doping regimes~\cite{zhu22}.
The results are shown in Fig.~\ref{fig:cooper}.
Quasi-long-range-order observed in previous DMRG calculations was
a hint to the existence of a superconducting phase in the 2D system~\cite{zhu22}.
Our simulations provide, for the first time,
clear evidence of {\it true} long-range order in the 2D triangular FH model.
Furthermore, such long-range correlations increase with the hole concentration.
A striking observation is that the chiral order parameter, although reduced, remains finite at the intermediate hole density ($n=5/6$),
$\chi = 0.22(1)$.
At the highest hole concentration, the chiral order is further suppressed.
At $n=5/6$, the triplet correlations at large distances are slightly stronger than the singlet ones.
Upon further doping the $2$D system,
the singlet and triplet correlations are (almost) degenerate.

We considered the effect of reduced dimensionality by computing the Cooper-pair correlations in our $36\times 3$ lattice. As expected, we do not see long-range order in quasi-$1$D.
Still, in the quasi-$1$D simulations, the triplet and singlet correlations showed to be degenerate (see Appendix~\ref{sec:1-2D}).
Notably, recent DMRG simulations showed a strong finite-size dependence. Indeed, singlet-pair correlations dominate for small system sizes, whereas triplet-pair correlations are enhanced for wider cylinders~\cite{zhu22}.

\begin{figure}[h]
\centering
\includegraphics[scale=0.8]{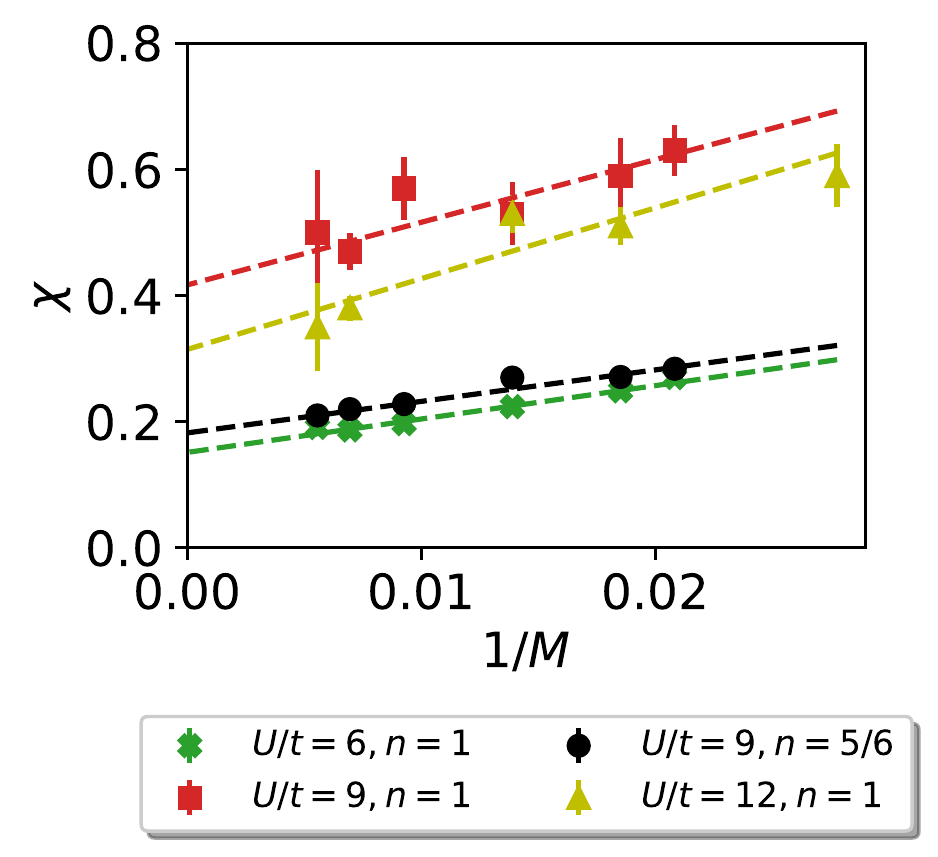}
\caption{{\bf Chiral order parameter} as a function of the inverted lattice size $1/M$.
The dashed lines are linear fits showing the extrapolation to the thermodynamic limit $1/M\rightarrow 0$.
\label{fig:extrap_chi}}
\end{figure}

The source of bias in our results is twofold: finite size and the CP approximation.
We analyzed the effect of finite size in the chiral order
parameter $\chi$ and performed extrapolation to the
thermodynamic limit in four situations where we
observed chirality: i) in the metallic phase at half filling ($U/t=6$);
ii) in the non-magnetic-insulating phase at half filling ($U/t=9$); iii) in the superconducting phase at filling $n = 5/6$ and interaction $U/t=9$; and
iv) in the AFM phase at half filling ($U/t = 12$).
For the first three cases, simulations were performed with lattice sizes $M=48$ ($N_x \times N_y = 6\times 8$), $M=54$ ($9\times 6$), $M=72$ ($6\times 12$), $M = 108$ ($9\times 12$),  $M = 144$ ($12\times 12$), and $M = 180$ ($12\times 15$).
The $M=48$ lattice under PBC is not compatible with the 120$^{\circ}$ AFM phase,
therefore, for this situation, we consider an $M = 36$ ($6 \times 6$) lattice.
Results are shown in FIG.~\ref{fig:extrap_chi}, where we see evidence that $\chi$ would remain finite when extrapolated to the thermodynamic limit $M\rightarrow\infty$. For the doped case, the simulation of the filling $n=5/6$ is not compatible with the $M=54$ lattice. Therefore, for this calculation, we used $n=22/27$.

To measure the bias from the CP approximation, we computed $\chi$ for the system described by the mean-field Hamiltonian of Eq.~\eqref{eq:mfh}. Our results for $U_{\text{eff}}/t=4$ is zero, showing that chirality for $U/t \geq 4$ emerges from the imaginary-time evolution regardless of the system size.

\section{Discussion and Conclusion.}
We implemented for the first time AFQMC simulations of the zero-temperature triangular-lattice FH model.
The method of choice is a state-of-the-art technique that provides unbiased estimates subjected
to a small systematic error arising from constraints that eliminate the fermionic sign problem.
We located the transition to the insulating phase with a critical interaction between $7 < U_{c1}/t\leq 8$.
The transition to the magnetic phase was located at $10 < U_{c2}/t \le 11$.
We observe a finite value of the chiral order parameter in all phases observed, and in the insulating phase there is competition between chiral and magnetic orders.
Finally, we considered hole doping the non-magnetic CSS near the phase transition at $U/t = 9$.
Our results show long-range order in the Cooper-pair correlations.

Topological states that break time-reversal symmetry
and have quasi-long-range order in the Cooper-pair correlations
were observed in DMRG simulations of the $t-J$ model extended to include three-spin chiral interactions~\cite{hua22}.
Here we reported the simulation of a chiral superconductor in the 2D triangular Hubbard model.
For filling $n=5/6$, we see a finite superconducting order parameter with chiral order and enhanced triplet-pair correlations.
A recent DMRG study also finds emergent chiral superconductivity in the $t-J$ model~\cite{hua23},
which further supports our results.

A natural followup of this work would include: ({\it i}) The characterization of
our CSS via spin-flux insertion to compute the Chern number ~\cite{sza20,hua22}; ({\it ii})
The evaluation of nonlocal correlations to uncover string patterns~\cite{chi19}; ({\it iii})
The investigation of quasi-periodic lattices~\cite{vie19,sbr20}
where the interplay between aperiodicity and long-range order leads to exotic phases, analogously to supersolidity~\cite{cot22} and glass physics~\cite{gau21,cia22} in Bose systems; ({\it iv}) The study of the extended FH model.
Long-range interactions can be engineered via Rydberg dressing~\cite{mac14,sch18,def21}, as recently realised in square lattices with unidirectional hopping~\cite{gua21}.

\begin{acknowledgments}
We acknowledge Peter Schauss, Davis Garwood, Natanael Costa, Bruno R. de Abreu, Silvio A. S. Vitiello and Federico Becca for insightful discussions.
This research was supported by High Performance Computing Center at UFRN (NPAD/UFRN) and Centro Nacional de Processamento
de Alto Desempenho em S\~{a}o Paulo (CENAPAD-SP) where the simulations where performed.
The software CPMC-Lab~\cite{ngu14} was used as an initial guide.
V. Z. acknowledges financial support
from the Brazilian agencies Coordena\c{c}\~{a}o de Aperfei\c{c}oamento de Pesquisa de Pessoal de
N\'{i}vel Superior (CAPES) under the Netherlands Universities Foundation for International Cooperation (NUFFIC)
exchange program (process number 88887.649143/2021-00) and the Serrapilheira Institute
(grant number Serra-1812-27802). T. M. acknowledges ITAMP for hospitality, where part of this work was done.
Our simulations were performed using release CP-AFQMC-v1.0 of the code available at the \href{https://github.com/quantechsimulations/CP-AFQMC}{Quantech simulations repository}~\cite{cp-afqmc_code}.
\end{acknowledgments}


\bibliography{fermhub}

\onecolumn\newpage
\appendix
\section{Constrained path AFQMC}\label{sec_mc}

Let a given initial state $|\Psi(0)\rangle$ be nonorthogonal to the ground-state of the Hamiltonian $H$, $|\Psi_0\rangle$.
The imaginary-time evolution $|\Psi(\tau)\rangle=\exp(-\tau H)|\Psi(0)\rangle$ will asymptotically converge to $|\Psi_0\rangle$
as $\tau$ (a real number) increases.
In the AFQMC method the anti-symmetric wave function is written as a linear combination of Slater determinants
\begin{equation}\label{eq_basis}
\left|\Psi(\tau)\right> = \sum_k\xi(\Phi_k)\left|\Phi_k(\tau)\right>.
\end{equation}
In our simulations, $\xi(\Phi_k)$
is not considered explicitly.
As the imaginary-time evolution proceeds, 
Slater determinants are replicated or killed.
The number of a given $|\Phi_k\rangle$ in the sum
resembles $\xi(\Phi_k)$~\cite{zha95,zha97}.
In practice one starts at $\tau = 0$ with all Slater Determinants equal to a given trial state $|\Phi_T\rangle$, an approximation to the ground state usually obtained from mean-field theories, and update them by the application of the imaginary-time evolution operator in a stochastic process.
For large systems the diagonalization of the FH Hamiltonian becomes impractical.
It is then usual to consider the Trotter formula~\cite{tro59} to factor the evolution operator into the product of three therms,
\begin{equation}\label{eq_tf}
e^{-\delta\tau H}=e^{-\frac{\delta\tau}{2}K}e^{-\delta\tau V}e^{-\frac{\delta\tau}{2}K} + {\cal O}(\delta\tau^2),
\end{equation}
where, for the Fermi Hubbard (FH) Hamiltonian, $K$ ($V$) is the hopping (interaction) therm.
If we chose $\delta\tau$ small enough, 
the error introduced by neglecting the ${\cal O}(\delta\tau^2)$ therms in Eq.~\eqref{eq_tf} can be made smaller than the statistical uncertainty inherent to Monte Carlo calculations;
the method remains numerically exact.
The desired limit $\tau = n\delta\tau \gg t^{-1}$, with $t$ being the hopping strength, is obtained after $n$ successive applications of this small-$\delta\tau$ approximation to $|\Psi(0)\rangle$. A given iteration on the Slater Determinants is
\begin{equation}\label{eq:it}
|\Phi^{n+1}_k\rangle = e^{-\frac{\delta\tau}{2}K}e^{-\delta\tau V}e^{-\frac{\delta\tau}{2}K} |\Phi^{n}_k\rangle,
\end{equation}
where the superscript $n$ indicates the imaginary time $\tau = n\delta\tau$.
The application of one-body operators in $|\Phi_k^n\rangle$ results in another Slater Determinant. Therefore,
both $\exp(-\delta\tau K/2)$ only propagates $|\Phi_k^n\rangle$.
On the other hand, since $V$ is a sum of two-body operators,
the remaining exponential imposes a challenge.
To handle it,
one can use the Hubbard-Stratonovich decomposition to transform the two-body interaction into one-body ones between each electron and auxiliary fields $x$. We choose the spin discrete decomposition~\cite{hir85}, 
\begin{equation}\label{eq_dhsd}
e^{-\delta\tau n_{i\uparrow}n_{i\downarrow}}=
e^{-\frac{\delta\tau}{2}U(n_{i\uparrow}+n_{i\downarrow})}
\displaystyle \sum_{x=\pm 1}p(x)
e^{\gamma x (n_{i\uparrow}-n_{i\downarrow})},
\end{equation}
with $p(x)=1/2$ and $\gamma$ being given by the relation $\cosh(\gamma)=\exp(\delta\tau U/2)$.
Considering the FH Hamiltonian $H$, 
one writes
\begin{equation}\label{eq_dm}
e^{\delta\tau H} \approx
\displaystyle\sum_{\vec{x}}p(\vec{x})
e^{-\frac{\beta}{2}K}
B_V(\vec{x})
e^{-\frac{\beta}{2}K},
\end{equation}
where
$\vec{x}=(x_1,x_2,...,x_M)$ is a configuration of auxiliary fields, with $M$ the number of lattice sites, to be sampled within Monte Carlo calculations,
$p(\vec{x})=(1/2)^M$ is a probability distribution function (pdf) and $B_V(\vec{x})$ is a product of one-body exponentials.
Explicitly,
\begin{equation}\label{eq_b}
B_V(\vec{x})=
\prod_i e^{-\frac{\delta\tau}{2}U(n_{i\uparrow}+n_{i\downarrow})+\gamma x_i (n_{i\uparrow}-n_{i\downarrow})}.
\end{equation}

The stochastic simulation of Eq.~\eqref{eq:it} is very inefficient since $p(\vec{x})$ is constant, and an importance sampling technique is required~\cite{zha95,zha97}.
The importance sampling is also useful to define an estimator to the properties of the system and to determine the constraints that eliminate the sign problem.
The importance function we implement is 
$O_T(\Phi_k^n)=\left<\Phi_T\right|\left.\Phi_k^n\right>$,
which leads to the modified imaginary-time evolution
\begin{equation}\label{eq:it_is}
|\tilde{\Psi}^{n+1}\rangle = \sum_{\vec{x}}\tilde{p}(\vec{x})e^{-\frac{\delta\tau}{2}K}B_V(\vec{x})e^{-\frac{\delta\tau}{2}K}|\tilde{\Psi}^n\rangle,
\end{equation}
with the modified pdf
$\tilde{p}(\vec{x})=O_T(\Phi_j^n)p(\vec{x})/O_T(\Phi_j^{n-1})$.
The pdf now depends on the overlap with the approximated ground state
after and before the diffusion procedure.
Since the pdf $\tilde{p}(\vec{x})$ is usually not normalized,
we define the normalization factor of each Slater determinant $N(\Phi_k^n)$,
and the iterative projection equation becomes
\begin{equation}\label{eq_is}
    \left|\Phi^n_k\right>=N(\Phi_k^n)\sum_{\vec{x}}
    \frac{\tilde{p}(\vec{x})}{N(\Phi_k^n)}e^{-\frac{\delta\tau}{2}K}B_Ve^{-\frac{\delta\tau}{2}K}
    \left|\Phi^{n-1}_k\right>.
\end{equation}
To handle the normalization we introduce weights to each Slater Determinant $|\tilde{\Psi^n}\rangle = \sum_k \omega_k^n|\Phi_k^n\rangle$,
and in each iteration the weights are updated as $\omega^n_k = N(\Phi_k^n)\omega^{n-1}_k$, with $\omega_k^0=1$ .
In practice, the sampling of the pdf $\tilde{p}(\vec{x})$ is done considering individually each auxiliary field in the configuration $\vec{x}$.
A detailed description of how to implement the aforementioned sampling and how to updated the weights is given in Ref.~\cite{ngu14}.

The equivalence between
$\left|\Phi_k^n\right>=-\left|\Phi_k^n\right>$
causes a sign problem that prevents numerical convergence.
To eliminate the sign problem,  auxiliary-field paths are constrained to a region of the configuration space were $O_T(\Phi_k^n) > 0$ (akin to the 
fixed-node approximation~\cite{rey82}).
Due to the equivalence between the positive and negative regions of the ground state,
the method is numerically exact if the nodal structure of the trial wave function is equivalent to the ground-state one.
Unfortunately the latter is unknown in the majority of cases and $\left|\Phi_T\right>$ is used as an approximation to it.
For that reason, the constrained-path approximation has a systematic error shown to be small~\cite{zha95,zha97,ngu14,qin16}.

Ground-state estimates of the system total energy can be obtained using the mixed estimator,
\begin{equation}\label{eq_mixed}
\langle H \rangle_{\text{mix}} =
\frac{\sum_k \omega_k^nE_k^n}{\sum_j\omega_k^n},
\end{equation}
with $E^n_k=\langle\Phi_T|H|\Phi^n_k\rangle/O_T(\Phi_k^n)$ and sufficiently large $n$.
As can be noticed,
the mixed estimator is only exact if the operator in the numerator of Eq.~\eqref{eq_mixed} commutes with the Hamiltonian $H$.

Estimates of other physical observables require the back-propagation technique~\cite{zha95,zha91,pur04}.
The back-propagation estimator is constructed from the formula

\begin{equation}\label{eq_bp}
\left< O\right>_{\text{bp}} \propto \langle \Phi_T|e^{-\tau_{\text{bp}}H}Oe^{-(\tau-\tau_{\text{bp}})H}|\Psi(0)\rangle,
\end{equation}
which asymptotically reaches the average value of the observable $O$ in the ground state for $\tau - \tau_{\text{bp}}$ and $\tau_{\text{bp}}\gg t^{-1}$.
The numerical evaluation of Eq.~\eqref{eq_bp} is implemented efficiently by storing the auxiliary fields sampled in the forward propagation
$\exp(-\tau_{\text{bp}}H)|\Psi(0)\rangle$ and using them to back propagate $\langle\Phi_T|$. For a detailed description of this estimator see Ref.~\cite{pur04}.

{\it Additional simulation details}.
Our time step is $\delta\tau = 10^{-2}/t$, with $10^3$ Slater determinants.
Also, we see convergence of our estimates with $\tau_{\text{bp}}=1.6/t$.

{\it Instability in the strong interaction regime}.
We noticed that for $U/t \ge 10$, the weights of the Slater Determinants suffer from strong oscillations, which cause higher variances in our final results.
These oscillations are controlled by the importance function, i.e., the mean-field {\it ansatz} being projected in imaginary time.

\begin{figure}[h]
\centering
\includegraphics[scale=0.8]{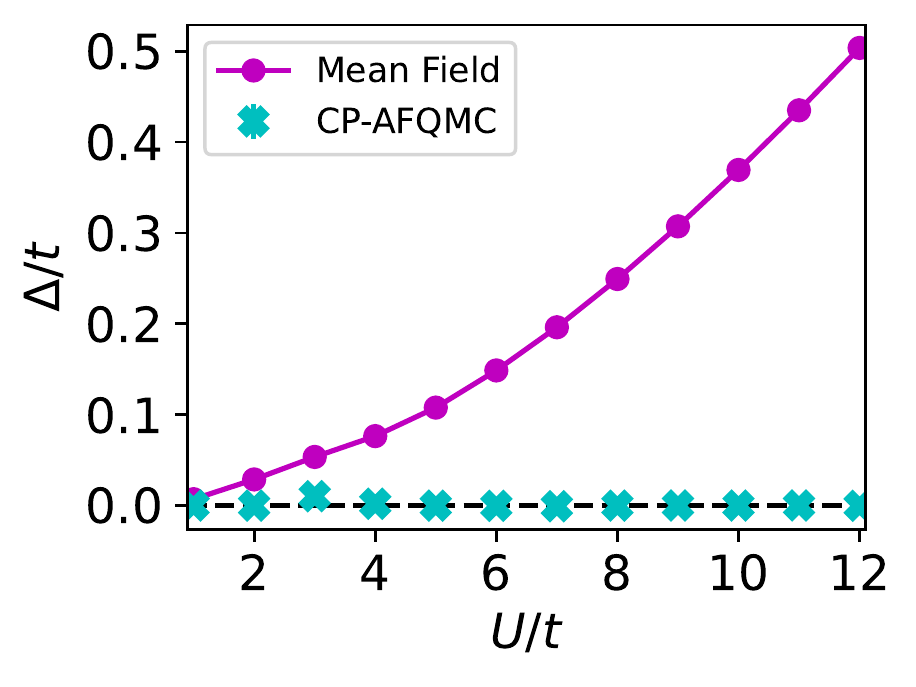}
\caption{{\bf Comparison with exact diagonalization.}
The relative difference between the energies is defined as $\Delta = (E - E_{\text{ED}})/E_{\text{ED}}$,
where $E_{\text{ED}}$ was computed with exact diagonalization and $E$ with the CP-AFQMC simulation or the mean-field ansatz. The mean-fielf ansatz is the one in Eq.~\eqref{eq:mfh} with $U_{\text{eff}}=\text{min}(U,4t)$.
\label{fig:ed}}
\end{figure}

{\it Comparison with exact diagonalization}.
To have an estimate of the effect of the CP approximation in our simulations, we considered a $3\times 3$ triangular lattice filled with $6$ electrons under open boundary conditions. We computed the relative difference between the ground-state energy calculated with exact diagonalization and CP-AFQMC. The results are shown in FIG.~\ref{fig:ed} where we also show the relative difference associated with the mean-field GHF ansatz with $U_{\text{eff}}=\text{min}(U,4t)$.

\newpage

\section{Mean-field ans\"{a}tze}\label{sec:mf}

The FHF Hamiltonian~\cite{cos18} that we test in the regime of strong interactions is
\begin{equation}\label{eq:fhf}
    H_{\text{FHF}}=-t\sum_{\langle ij\rangle\sigma}(c^{\dagger}_{i\sigma}c_{j\sigma} + \text{H.c.})
    +U_{\text{eff}}\sum_i \left(\frac{1}{2}\langle n_i\rangle n_i - \frac{1}{4}\langle n_i\rangle^2
    - \frac{1}{2}\langle {\bf S}_i\rangle \cdot {\bf S}_i + \frac{1}{4}\left|\langle {\bf S}_i\rangle\right|^2\right),
\end{equation}
where we omitted the terms that do not conserve the number of particles.
We tested the FHF ansatz and the GHF one, from Eq.~\eqref{eq:mfh}, considering $U_{\text{eff}}=\text{min}(U,U_{\text{eff}}^{\text{max}})$ with $U_{\text{eff}}^{\text{max}}=2t$, $4t$ and $6t$. The results for the ground-state energy are in Table \ref{tab:mf}.

\begin{table}[h]
    \centering
    \begin{tabular}{c|c|c|c}
        \hline \hline
            $U_{\text{eff}}^{\text{max}}$ & $2t$ & $4t$ & $6t$ \\
        \hline
        GHF ($U = 9t$)  & $E/t = -77.0(1)$ & $E/t = -76.9(1)$ & $E/t = -76.9(1)$ \\
        FHF ($U = 9t$)  & $E/t = -77.1(1)$ & $E/t = -77.3(2)$ & $E/t = -73.03(8)$ \\
        \hline
        GHF ($U = 12t$) & $E/t = -53.5(3)$ & $E/t = -54.3(3)$ & $E/t = -54.3(3)$ \\
        FHF ($U = 12t$) & $E/t = -53.4(4)$ & $E/t = -54.3(4)$ & $E/t = -55.48(8)$ \\
        \hline \hline
    \end{tabular}
    \caption{{\bf Ground-state energy}. Simulations of the $12\times 12$ system with $U/t = 9$ ans $12$ and different initial ans\"{a}tze.}
    \label{tab:mf}
\end{table}

\begin{figure}[h]
\centering
\includegraphics[scale=0.7]{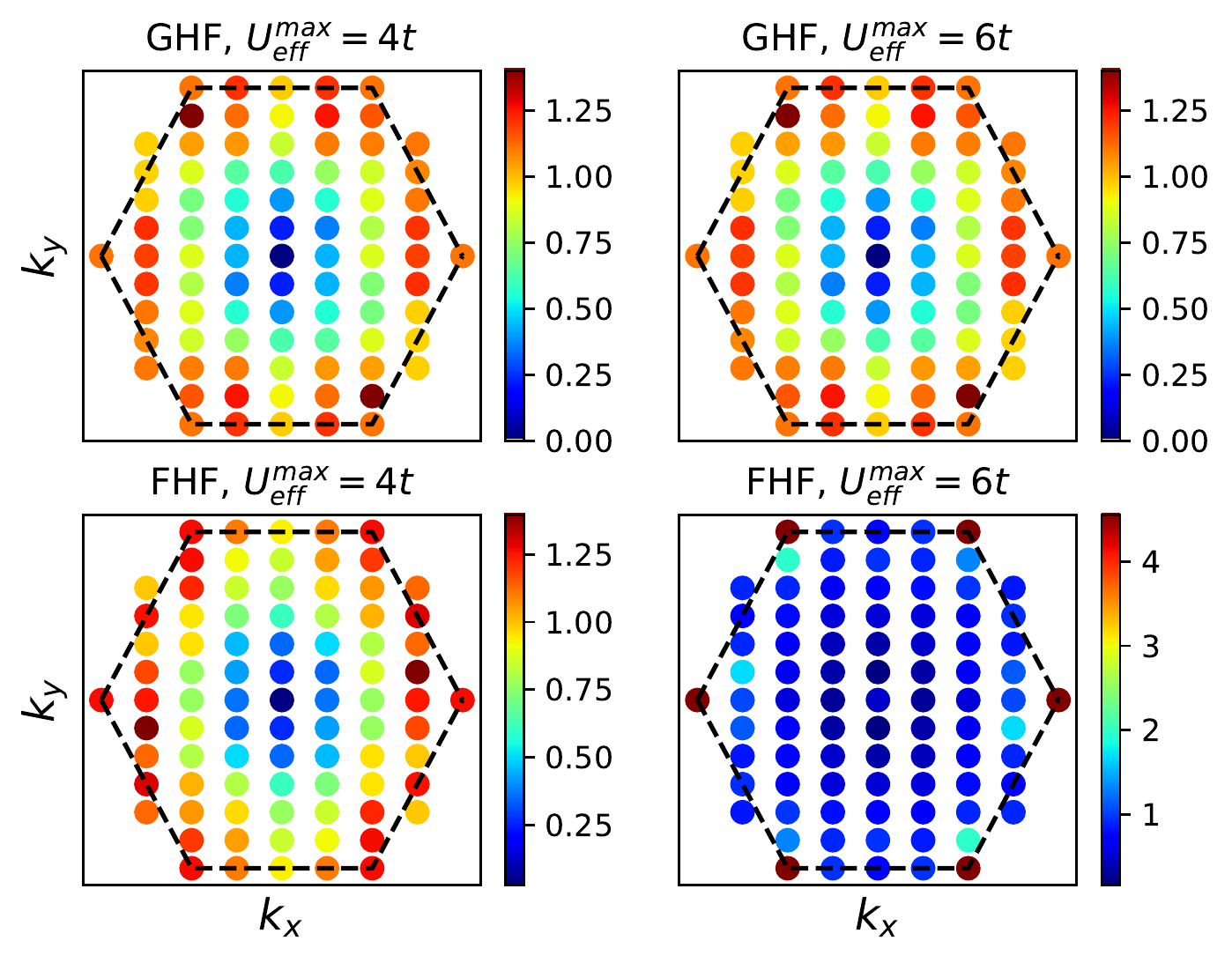}
\caption{{\bf Spin structure factor} for $U/t = 9$ after imaginary-time evolution from four different ans\"{a}tze.
All results are for allowed momenta in the considered geometry.
\label{fig:hf_u9}}
\end{figure}

Considering the variational principle,
the state that provides the smaller energy after imaginary-time evolution is the best representation
of the ground state of the system for a given value of $U/t$.
For example, in FIG.~\ref{fig:hf_u9} we show the spin structure factor of the $U/t = 9$ case after
imaginary-time evolution starting from four different ans\"{a}tze.
We see sharp peaks in the spin structure factor after evolution from the FHF ansatz with $U_{\text{eff}}^{\text{max}} = 6t$. Those peaks are evidence of long-range magnetic order, but since the other ans\"{a}tze provided a smaller energy,
the ground state of the system for $U/t = 9$ is non-magnetic.

\begin{figure}[h]
\centering
\includegraphics[scale=0.8]{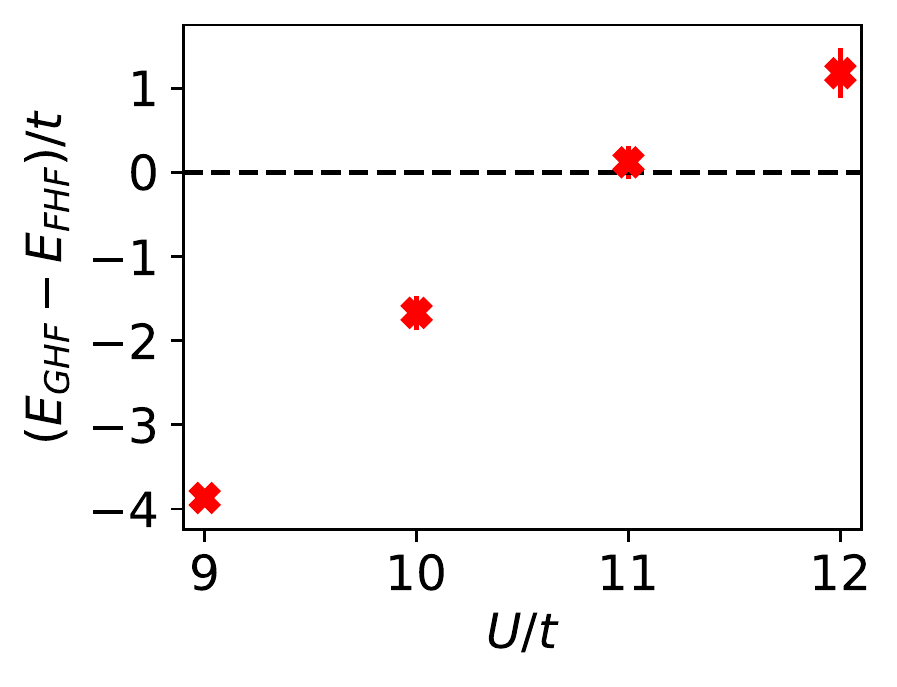}
\caption{Difference between energies obtained evolving the GHF ($U_{\text{eff}}^{\text{max}} = 4t$) and the FHF ($U_{\text{eff}}^{\text{max}} = 6t$) ans\"{a}tze in imaginary time. 
\label{fig:gxf}}
\end{figure}

We computed the spin structure factor of the FHF ansatz with $U_{\text{eff}}^{\text{max}} = 6t$ itself and concluded that long-range magnetic order after imaginary-time evolution is reminiscent of this initial choice.
Therefore, for $U/t \ge 9$ we perform simulations with two different ans\"{a}tze: a) the GHF one with $U_{\text{eff}}^{\text{max}} = 4t$ and b) the FHF one with $U_{\text{eff}}^{\text{max}} = 6t$.
After imaginary-time evolution, the state with smaller energy is our best representation of the ground state of the system for a particular value of $U/t$.
In FIG.~\ref{fig:gxf}, we show the difference between those energies, which indicates a phase transition to the magnetic phase with critical interaction close to $U/t = 11.$
We emphasize that, due to the CP approximation, our imaginary-time evolution is not variational. Therefore, care must be taken when comparing the ground-state energies projected from GHF and FHF ans\"{a}tze. Even though, we notice that the difference between those energies is always greater than our measure of the error due to the CP approximation (FIG.~\ref{fig:ed}), excluding the case of interaction $U/t = 11$. This corroborates the transition to the magnetic phase at strong interactions.

\newpage

\section{Finite system size effects}\label{sec:finsize}

\begin{figure}[h]
\centering
\includegraphics[scale=0.8]{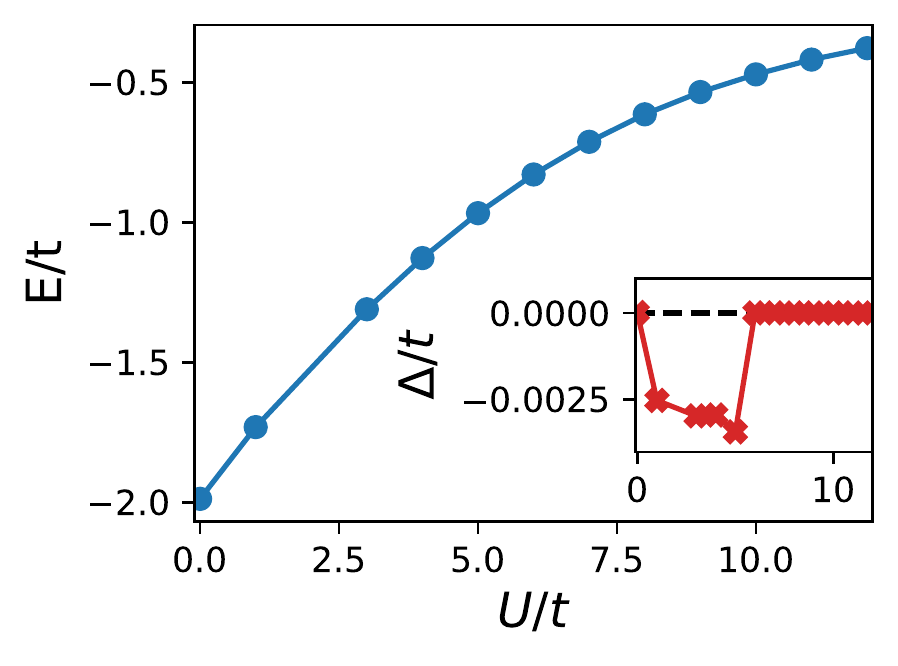}
\caption{{\bf Total energy per lattice site.}
The blue circles represent the $12\times 12$ lattice.
In the inset, we show the relative difference $\Delta = (E_{12\times 12}-E_{9\times 12})/E_{12\times 12}$. 
\label{fig:energy}}
\end{figure}

We compute the total energy of the system at half filling for two lattice sizes,
with $108$ sites ($N_x \times N_y = 9 \times 12$) and $144$ sites ($N_x \times N_y = 12 \times 12$).
The energies are shown in FIG.~\ref{fig:energy}.

\begin{figure}[h]
\centering
\includegraphics[scale=0.8]{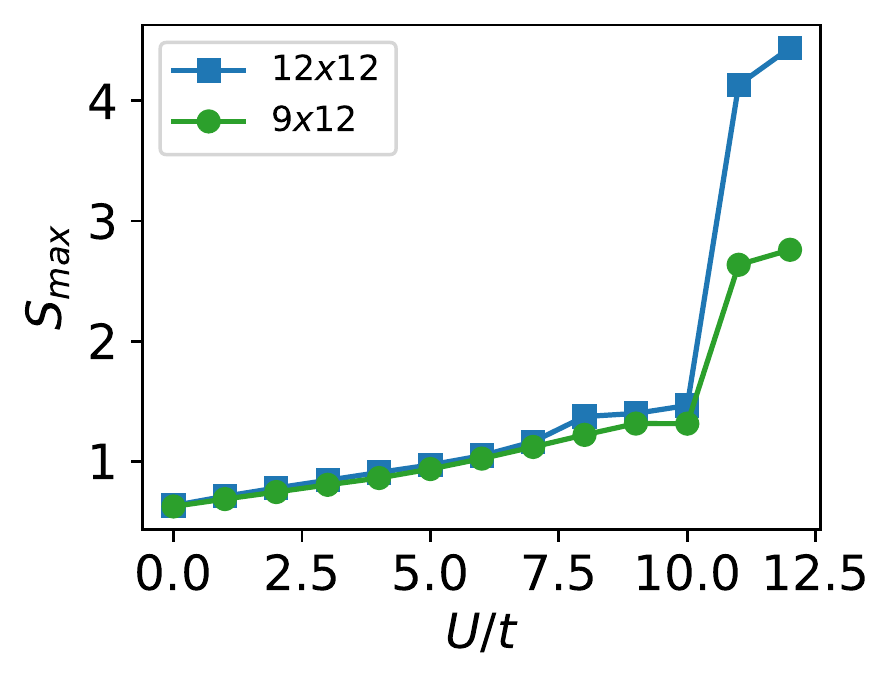}
\caption{{\bf Spin structure factor}.
The maximum value of $S(k)$ for two different lattice sizes.
\label{fig:magxmag}}
\end{figure}

We compute $S_{\text{max}}$ for the $12\times 12$ and $9 \times 12$ lattices to see the effect of the system size in the spin correlations.
Results are shown in FIG.~\ref{fig:magxmag}, where size effects influence the spin correlations from intermediate to strong coupling.

\newpage

\section{Comparison between 2D and quasi-1D}\label{sec:1-2D}

\begin{figure}[h]
\centering
\includegraphics[scale=0.8]{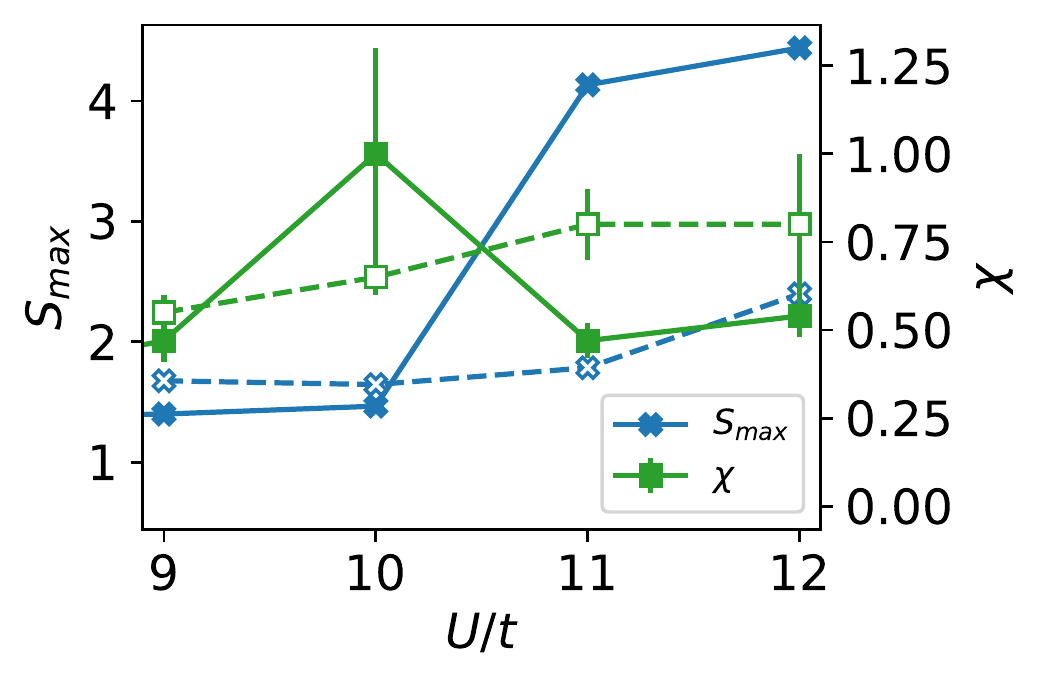}
\caption{(Color online) {\bf Magnetic versus chiral orders} at half filling. Comparison between results for the 2D lattice ($12\times 12$) and the quasi-1D ($36\times 3$) one as function of  $U/t$.
Filled (empty) markers are results for the 2D (quasi-1D) lattice.
\label{fig:magvschi36x3}}
\end{figure}

In FIG.~\ref{fig:magvschi36x3} we compare results for $S_{\text{max}}$ and $\chi$ in 2D ($12\times 12$) and quasi-1D ($36 \times 3$). We see that, for the quasi-1D system, $S_{\text{max}}$ suddenly increases for interactions between $11 < U/t \leq 12$ indicating the transition to the AFM phase. For the 2D lattice, the increment of $S_{\text{max}}$ happens earlier and is much more pronounced. We emphasize that all simulations in quasi-1D were made considering the GHF ansatz.

\begin{figure}[h]
\centering
\includegraphics[scale=0.7]{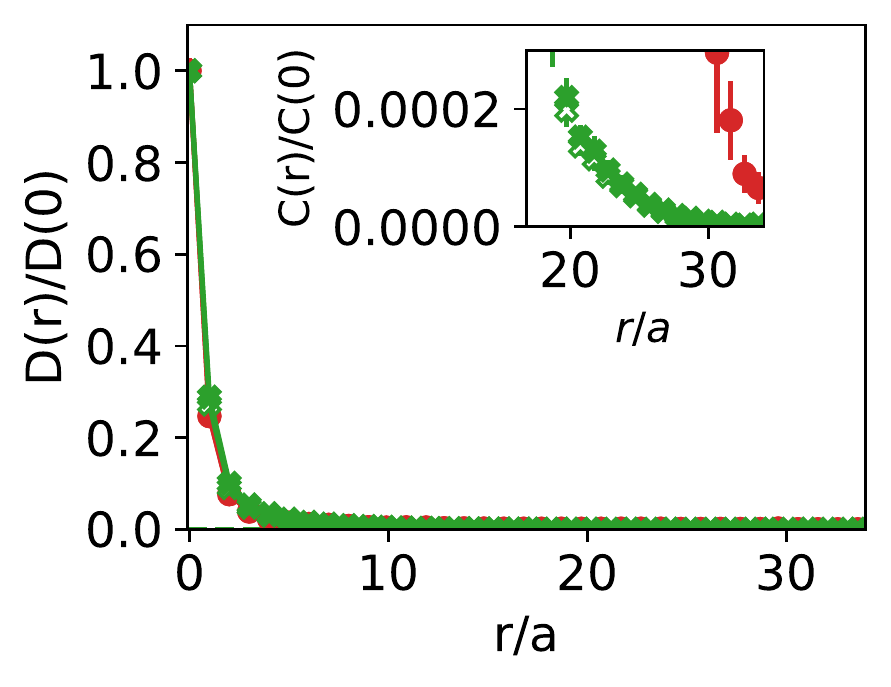}
\caption{(Color online) {\bf Cooper pair correlations.} Results for the quasi-1D ($36\times 3$) lattice. Green crosses represent the results at $n = 5/6$, and red circles are results at $n = 17/18$. Filled symbols are for triplet pairs, and empty ones for singlets (they are degenerate). The inset shows the long-range behavior. The interaction is $U/t = 9$.
\label{fig:cooper36x3}}
\end{figure}

We also compute the Cooper pair correlations for the quasi-1D system at $U/t = 9$, and as expected by the Mermin-Wagner theorem, they do not display long-range order. We considered the fillings $n = 17/18$ and $n = 5/6$, see FIG.~\ref{fig:cooper36x3}.

\begin{figure}[h]
\centering
\begin{subfigure}{.5\textwidth}
  \centering
  \includegraphics[width=1\linewidth]{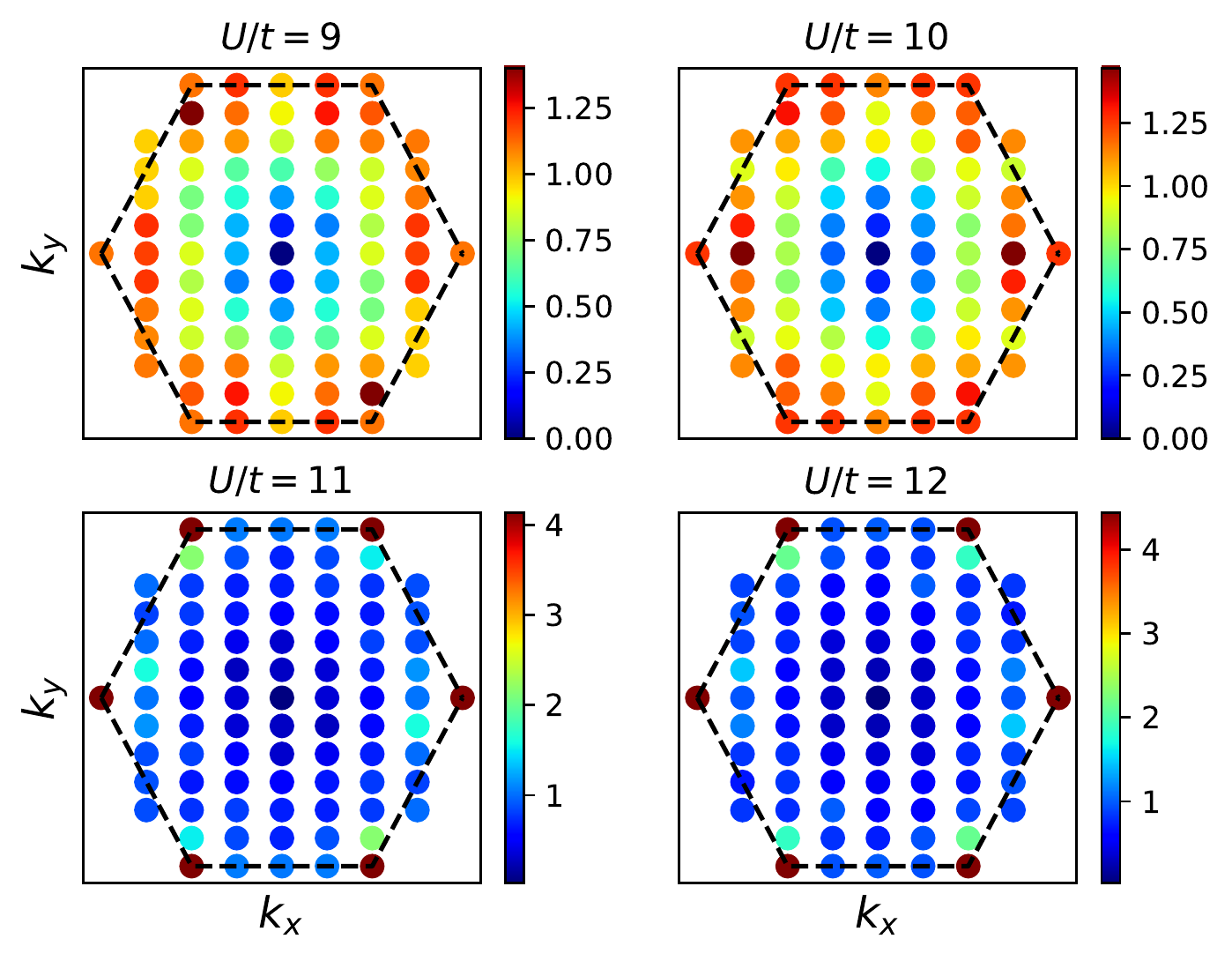}
  \caption{}
  \label{fig:sub1}
\end{subfigure}%
\begin{subfigure}{.5\textwidth}
  \centering
  \includegraphics[width=1\linewidth]{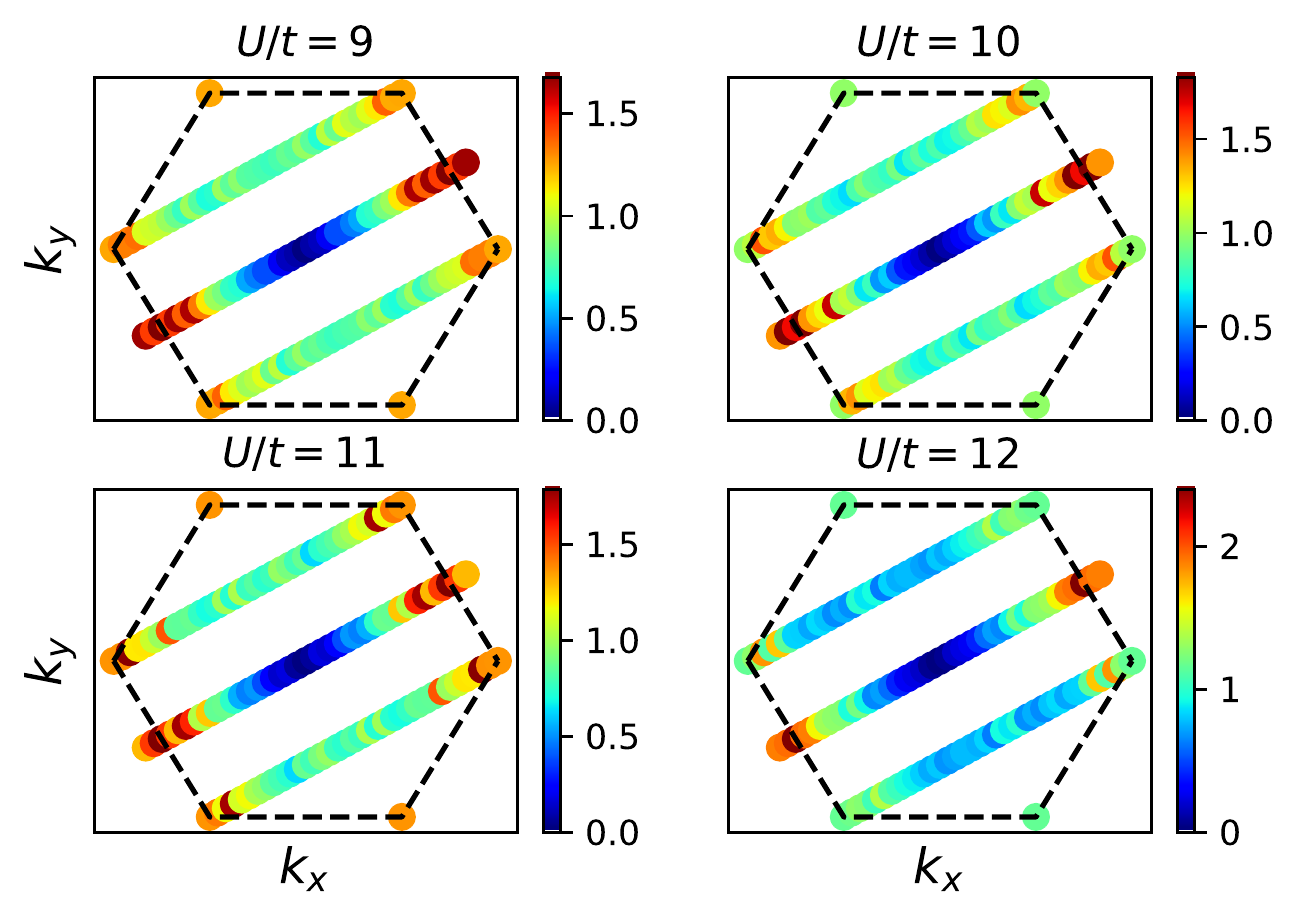}
  \caption{}
  \label{fig:sub2}
\end{subfigure}
\caption{{\bf Spin structure factor} for interactions $U/t = 9,10,11$ and $12$ at half filling.
(a) Results for the 2D lattice with $144$ sites. (b) Results for the quasi-1D system with $108$ sites.
The dashed hexagons in (a) and (b) delimit the first Brillouin zone of the triangular lattice.
$K$ points are on the vertexes of the hexagons, and $M$ points are in the middle of its edges.
All results are for allowed momenta in the considered geometry.}
\label{fig:ssf}
\end{figure}

Furthermore, In FIG.~\ref{fig:ssf} we show the spin structure factor $S(k)$ for $U/t=9,10,11$ and $12$.
The maxima of $S(k)$ is around $K$ for the 2D lattice, but it moves to the $M$ in quasi-1D.
In both cases one can see the formation of peaks indicating spin order.

\begin{figure}[h]
\centering
\begin{subfigure}{.5\textwidth}
  \centering
  \includegraphics[width=1\linewidth]{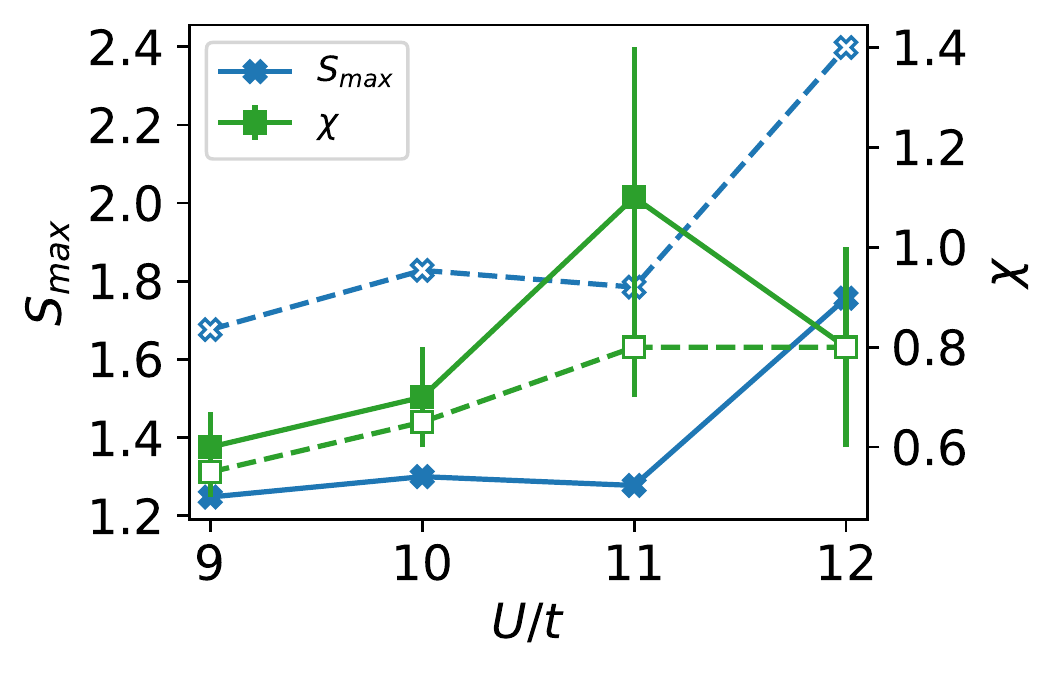}
  \caption{}
  \label{fig:sub11d}
\end{subfigure}%
\begin{subfigure}{.5\textwidth}
  \centering
  \includegraphics[width=1\linewidth]{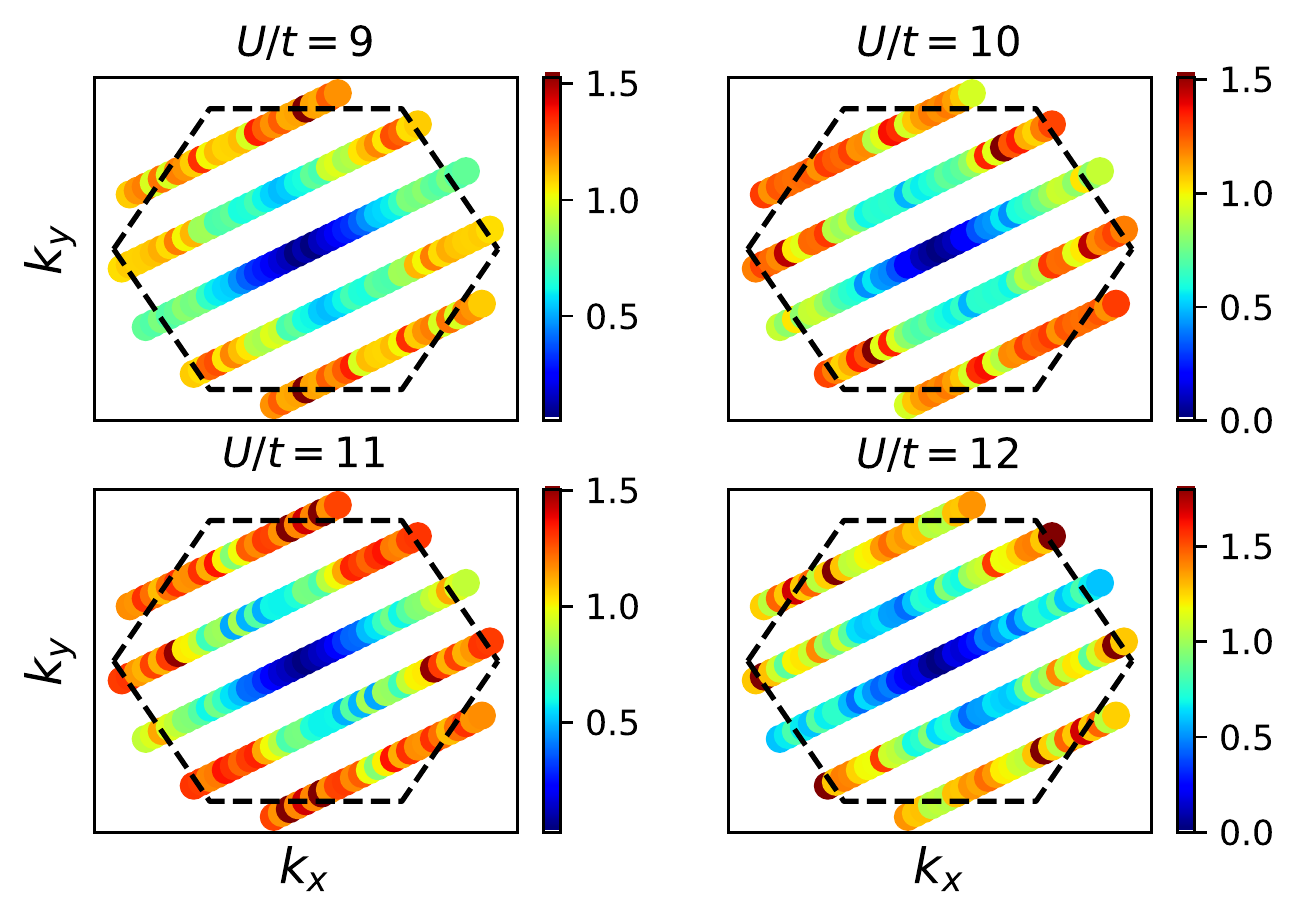}
  \caption{}
  \label{fig:sub21d}
\end{subfigure}
\caption{{\bf Spin structure factor and chirality} for interactions $U/t = 9,10,11$ and $12$ at half filling in quasi-1D.
(a) Comparison between the results of the $3\times 36$ (empty markers) and the $4\times 36$ (full markers) systems.
(b) Spin structure factor of the $4\times 36$ lattice.}
\label{fig:ssf1d}
\end{figure}

Finally, we analyze the influence of the width of the quasi-1D systems in our results. We compare the chirality and the spin correlations in $36\times 3$ and $36 \times 4$ lattices; results are shown in FIG.~\ref{fig:ssf1d}. In both systems, we see the transition to the magnetic phase marked by a sudden increase of $S_{\text{max}}$, but for the quasi-1D cylinder of width four, the peaks of the spin structure factor $S(k)$ seem to be more intense in the $K$ rather than in the $M$ points as was noticed in the width-three system (refer to FIG.~\ref{fig:sub2} for the width-three result). Nonetheless, we also see high values of $S(k)$ at the $M$ points for the width-four system. This indicates competition between two different magnetic orders as reported in Ref.~\cite{wie21}. Chirality is also seen in both systems.

\end{document}